\documentstyle[]{mn}

\def\eps@scaling{.95}
\def\epsscale#1{\gdef\eps@scaling{#1}}

\def\plotone#1{\centering \leavevmode
    \epsfxsize=\eps@scaling\columnwidth \epsfbox{#1}}

\def\plotfiddle#1#2#3#4#5#6#7{\centering \leavevmode
    \vbox to#2{\rule{0pt}{#2}}
    \includegraphics{#1}}

\newif\ifAMStwofonts
\ifoldfss
  \ifCUPmtlplainloaded \else
    \NewTextAlphabet{textbfit} {cmbxti10} {}
    \NewTextAlphabet{textbfss} {cmssbx10} {}
    \NewMathAlphabet{mathbfit} {cmbxti10} {} 
    \NewMathAlphabet{mathbfss} {cmssbx10} {} 
  \fi
  \ifAMStwofonts
    \ifCUPmtlplainloaded \else
      \NewSymbolFont{upmath} {eurm10}
      \NewSymbolFont{AMSa} {msam10}
      \NewMathSymbol{\upi}     {0}{upmath}{19}
      \NewMathSymbol{\umu}     {0}{upmath}{16}
      \NewMathSymbol{\upartial}{0}{upmath}{40}
      \NewMathSymbol{\leqslant}{3}{AMSa}{36}
      \NewMathSymbol{\geqslant}{3}{AMSa}{3E}

    \fi
  \fi
\fi 

\ifnfssone
  \newmathalphabet{\mathit}
  \addtoversion{normal}{\mathit}{cmr}{m}{it}
  \addtoversion{bold}{\mathit}{cmr}{bx}{it}
  \newmathalphabet{\mathbfit} 
  \addtoversion{normal}{\mathbfit}{cmr}{bx}{it}
  \addtoversion{bold}{\mathbfit}{cmr}{bx}{it}
  \newmathalphabet{\mathbfss} 
  \addtoversion{normal}{\mathbfss}{cmss}{bx}{n}
  \addtoversion{bold}{\mathbfss}{cmss}{bx}{n}
  \ifAMStwofonts
    \ifCUPmtlplainloaded \else
      \UseAMStwoboldmath
      \makeatletter
      \new@mathgroup\upmath@group
      \define@mathgroup\mv@normal\upmath@group{eur}{m}{n}
      \define@mathgroup\mv@bold\upmath@group{eur}{b}{n}
      \edef\UPM{\hexnumber\upmath@group}
      \new@mathgroup\amsa@group
      \define@mathgroup\mv@normal\amsa@group{msa}{m}{n}
      \define@mathgroup\mv@bold\amsa@group{msa}{m}{n}
      \edef\AMSa{\hexnumber\amsa@group}
      \makeatother
      \mathchardef\upi="0\UPM19
      \mathchardef\umu="0\UPM16
      \mathchardef\upartial="0\UPM40
      \mathchardef\leqslant="3\AMSa36
      \mathchardef\geqslant="3\AMSa3E
    \fi
  \fi
\fi 

\ifnfsstwo
  \DeclareMathAlphabet{\mathbfit}{OT1}{cmr}{bx}{it}
  \SetMathAlphabet\mathbfit{bold}{OT1}{cmr}{bx}{it}
  \DeclareMathAlphabet{\mathbfss}{OT1}{cmss}{bx}{n}
  \SetMathAlphabet\mathbfss{bold}{OT1}{cmss}{bx}{n}
  \ifAMStwofonts
    \ifCUPmtlplainloaded \else
      \DeclareSymbolFont{UPM}{U}{eur}{m}{n}
      \SetSymbolFont{UPM}{bold}{U}{eur}{b}{n}
      \DeclareSymbolFont{AMSa}{U}{msa}{m}{n}
      \DeclareMathSymbol{\upi}{0}{UPM}{"19}
      \DeclareMathSymbol{\umu}{0}{UPM}{"16}
      \DeclareMathSymbol{\upartial}{0}{UPM}{"40}
      \DeclareMathSymbol{\leqslant}{3}{AMSa}{"36}
      \DeclareMathSymbol{\geqslant}{3}{AMSa}{"3E}
    \fi
  \fi
\fi 

\ifCUPmtlplainloaded \else
  \ifAMStwofonts \else 
    \def\upi{\pi}
    \def\umu{\mu}
    \def\upartial{\partial}
  \fi
\fi

\title{The percentage of stellar light re-radiated by dust in late-type 
Virgo Cluster galaxies}

\author[C.C. Popescu and R.J. Tuffs]
{Cristina C. Popescu$^{1,2,4}$ and Richard. J. Tuffs$^{3}$\\
$^{1}$The Observatories of the Carnegie Institution of Washington,
813 Santa Barbara Str., Pasadena, 91101 CA, USA, popescu@ociw.edu\\
$^{2}$IPAC (Caltech/JPL), 770 S. Wilson Avenue, Pasadena, California 91125, 
USA\\
$^{3}$Max Planck Institut f\"ur Kernphysik, Saupfercheckweg 1, 
69117 Heidelberg, Germany, Richard.Tuffs@mpi-hd.mpg.de\\
$^{4}$Research Associate, The Astronomical Institute of the 
Romanian Academy, Str. Cu\c titul de Argint 5, Bucharest, Romania}

\date{Accepted 2002 July 19.
      Received 2002 May 17;}
\pagerange{\pageref{firstpage}--\pageref{lastpage}}
\pubyear{2002}

\begin{document}
\maketitle
\label{firstpage}

\begin{abstract}
We show that the mean percentage of stellar light
re-radiated by dust is $\sim 30\%$ for the Virgo Cluster 
late-spirals measured with ISOPHOT by Tuffs et al. (2002). A strong 
dependence of this ratio with morphological type was found, ranging from 
typical values of $\sim 15\%$ for early spirals to up to $\sim 50\%$ for some 
late spirals. The extreme BCDs can have even higher percentages of their 
bolometric output re-radiated in the thermal infrared. Luminosity correction
factors for the cold dust component are given for general use in converting
far-infrared (FIR) luminosities derived from IRAS.
\end{abstract}


\section{Introduction}

Star-forming galaxies contain dust which absorbs some fraction of the emitted 
starlight, primarily re-radiating it in the far-infrared (FIR). 
This not only applies to starburst and ultraluminous
systems, which radiate almost all their power in the FIR, but also to so-called
``normal'' galaxies - systems which are not dominated by AGN and not undergoing
a starburst. Though less spectacular than starburst galaxies, normal 
galaxies still account for most of the infrared emissivity of the local 
universe. Their role in the distant universe is observationally a 
completely open question.

Recent observations with the ISOPHOT instrument (Lemke et al. 1996) on board 
the Infrared Space Observatory (ISO; Kessler et al. 1996), covering the 
spectral peak of the dust emission between 100 and 200$\,{\mu}{\rm m}$, have 
shown that a significant contribution to the FIR luminosity of normal 
galaxies is actually radiated by grains too cold to be visible to
IRAS. Statistical evidence for the existence of a cold dust
component was established in our study (Popescu et al. 2002) of the spatially 
integrated FIR emissions of a complete volume- and luminosity sample of 63 
gas-rich Virgo Cluster galaxies measured with ISOPHOT at 60, 100 and 
170\,${\mu}$m (Tuffs et al. 2002).\footnote{A subsample of these galaxies was
also observed with the ISO LWS instrument by Leech et al. (1999).} These 
observations represent the deepest 
survey (both in luminosity and surface brightness terms) of normal galaxies 
yet measured in 
the FIR. In particular, these data showed that the cold dust component is 
present in all galaxies later than S0, i.e. spiral, irregular and blue compact 
dwarf (BCD) galaxies. 

In this letter, these new results are used to evaluate
the fraction of starlight emitted in normal galaxies
which is re-radiated in the FIR, the first such measurement
for these systems. Previous estimates based on the
IRAS Bright Galaxy Sample (BGS; Soifer \& Neugebauer 1991)
have established a canonical value of 30\% for the
fraction of starlight to be re-radiated in the FIR in
the local universe. However, this value refers to relatively bright
FIR sources in which the bulk of the dust emission is radiated in
the IRAS 60 and 100\,${\mu}$m bands, and is not representative
of quiescent systems like the Virgo galaxies.
In addition it takes no account of measurements longwards of 120\,${\mu}$m, not
available at that time. The percentage of stellar light re-radiated by dust was
investigated by Xu \& Buat (1995), using an indirect estimate for the total FIR
luminosity. Here we calculate this percentage by using for the first time
measurements of the bulk of the dust emission in quiescent normal
galaxies.

\begin{figure}
\plotfiddle{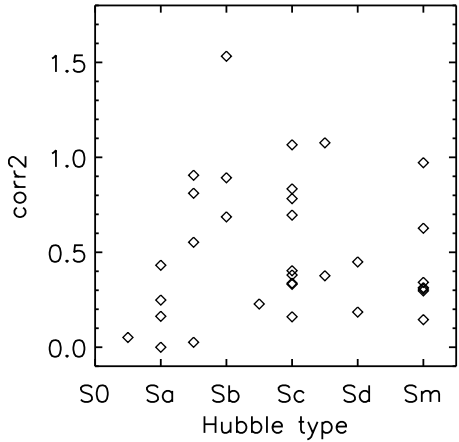}{2.2in}{0.}{60.}{60.}{-150}{-225}
\caption{The luminosity correction factor $corr2$ for the cold dust versus Hubble
type.}
\end{figure}

Before evaluating the percentage of stellar light re-radiated by dust (Sect.~4)
we describe our derivation of the total bolometric output in the ultraviolet 
(UV)/optical/near-infrared (NIR) in Sect.~3 and determine in Sect~2 luminosity 
correction factors for the cold dust, for general use in converting IRAS
luminosities into total FIR luminosities.

\section{Luminosity correction factors for the cold dust}

As shown by Popescu et al. (2002) the FIR spectral energy distribution (SED) from
normal galaxies in the 60 - 170\,${\mu}$m range is typically rather broad,
requiring warm and cold emission components. Especially for later spirals
and irregulars, most of
the FIR luminosity is carried by the cold dust component primarily
emitting longwards of the IRAS limit of 120\,${\mu}$m. Studies based
on IRAS data have used various corrections to account for the
emission beyond 120\,${\mu}$m, under the assumption of a single
dust temperature component (e.g. Helou et al. 1988).
In the light of our new results, these will underestimate the true FIR
luminosity, especially for normal galaxies.
Since most galaxies do not yet have measurements
at these longer wavelengths, it is useful to derive correction
factors based on our ISOPHOT Virgo cluster sample. Such corrections should be
generally applicable to convert IRAS luminosities into total FIR luminosities.

First we define a correction factor $corr1$
by which the total FIR luminosity $L_{\rm FIR}$ differs from the FIR luminosity 
$L_{40-120}$: 

\begin{eqnarray}
corr1 = \frac{L_{\rm FIR}-L_{40-120}}{L_{40-120}}
\end{eqnarray}\\
Here $L_{40-120}[{\rm W}/{\rm m}^2]=1.26\times
10^{-14}(2.58f_{60{\mu}m}+f_{100{\mu}m})$, where
$f_{60{\mu}m}$, $f_{100{\mu}m}$ are in Jy (Helou et al. 1988).
The total FIR luminosities $L_{\rm FIR}$ were taken from Table~2 and 3 
of Popescu et al. (2002). $L_{\rm FIR}$ were derived by fitting the FIR SEDs
with two modified blackbody (Planck) functions of fixed emissivity index
${\beta}=2$ (for details see Popescu et al. 2002) and integrating between 40 
and 1000\,${\mu}$m. Only galaxies with 
detections at all three wavelengths (60, 100 and 170\,$\mu$m) were 
considered, which make a total of 38 objects. 

We also define a correction factor $corr2$ by which $L_{\rm FIR}$
differs from a single dust component extrapolated total FIR luminosity 
$L_{60;100}^{\rm extrapolated}$:

\begin{eqnarray}
corr2 = \frac{L_{\rm FIR}-L_{60;100}^{\rm extrapolated}}{L_{60;100}^{\rm extrapolated}}
\end{eqnarray}\\
$L_{60;100}^{\rm extrapolated}$ were
calculated based solely on our 
ISOPHOT 60 and 100\,${\mu}$m flux densities, by 
fitting the FIR SED with a single modified blackbody (Planck) function, otherwise  
following the same procedure used to derive the total $L_{\rm FIR}$.

The results for the cold dust luminosity correction factors (only for $corr2$)
are illustrated in Fig.~1, as a function of Hubble type. A large scatter can
be seen, showing that it is not possible to apply a meaningful correction for the cold
dust to individual galaxies. Nevertheless, these numbers give some idea of the 
corrections to be expected for statistical samples.

\begin{table}
\caption{The cold dust luminosity correction factors (median, mean and standard
deviation of the mean)}
\begin{tabular}{ccccccc}
\hline\hline
Type  & \multicolumn{3}{c}{$corr1$} & \multicolumn{3}{c}{$corr2$} \\
       &median & mean & ${\epsilon}$ &median & mean & ${\epsilon}$ \\
\hline
S0a/Sa & 0.85 & 0.73 & 0.17 & 0.11 & 0.12 & 0.09\\
Sab-Sm & 1.04 & 1.11 & 0.09 & 0.43 & 0.56 & 0.07 \\
Im-BCD & 1.06 & 1.81 & 0.89 & 0.85 & 1.19 & 0.64 \\
\hline
\end{tabular}
\end{table}

Despite the scatter in Fig.~1, there is a visible trend for the 
early spirals (S0a, Sa) to have smaller corrections than the later
spirals. 
It is therefore meaningful 
to derive statistical corrections for the early spirals and the later spirals 
separately. The median (and mean)
values given in Table~1 indeed show a segregation between the early
and the later types. Not shown in Fig.~1 are the corrections for the
Im-BCDs. Since most of the BCDs were not detected at 60 and 100\,${\mu}$m
(Tuffs et al. 2002), suggesting the coldest dust temperatures from all 
Hubble types, only 3 galaxies could be used for this statistic. 
Their median (and mean) correction factor has the highest value. Possible 
explanations for the unusual behaviour of the 
Virgo BCDs were given in Popescu et al. (2002).

The correction factors for the cold dust luminosity were derived under the
assumption of an emissivity index $\beta=2$, which is consistent with the
graphite/silicate dust model of Draine \& Lee (1984). If we were to adopt an
emissivity index $\beta=1$, the correction factors for the cold dust luminosity
would decrease by $\sim 0.11$.

No evidence was found for a dependence of the luminosity correction factor on
galaxy mass, as traced by the dependence of the K$^{\prime}$ magnitudes from
Boselli et al. (1997).
  
\section{The UV/optical/NIR observed output}

In order to derive the total observed output in the optical bands, we
considered a subsample (28 galaxies) of our Virgo sample which had 
multi-aperture photometry in the Cousins U,B,V,R,I bands from 
Schr\"oder \& Visvanathan (1996; SV96). We applied the U-B, B-V, V-R and 
R-I colours taken from the largest aperture to the total $B_{\rm T}$ 
magnitude from Binggeli, Sandage \& Tammann (1985; BST85). This 
converts the aperture photometry to the total magnitudes, but is still affected
 by differences between the photographic system and the photometric 
Cousins system. To correct for the latter effect we compared the B
magnitudes taken from SV96 with those from BST85, for the galaxies 
having Binggeli's 25.5 B-mag\,arcsec$^{-2}$ isophotal diameters smaller 
or equal to the largest 142.7 arcsec aperture of SV96. We
found an average value $B_{\rm T}-B=-0.088$ which we applied to our derived
magnitudes. For the extinction in our Galaxy we used the $A_{\rm B}$ 
values from Burnstein \& Heiles (1982) and the $R_{\lambda}/R_{\rm V}$ 
factors from Cardelli, Clayton \& Mathis (1989). The conversion factors from 
magnitudes to flux densities were calculated from the Vega spectrum taken 
from Table~4 of Dreiling \& Bell (1980), and using the filter profiles 
from Table~2 of Bessell (1990). A white spectrum source was assumed.  

\begin{figure}
\plotfiddle{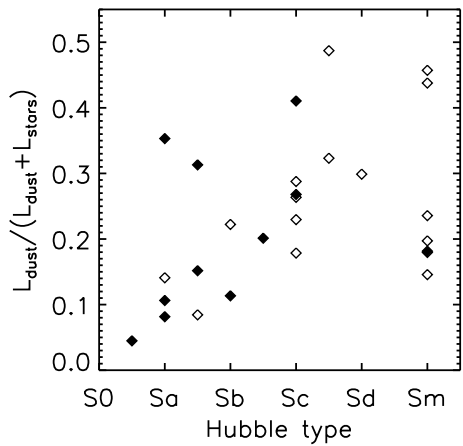}{2.2in}{0.}{60.}{60.}{-150}{-225}
\caption{The ratio of dust luminosity ($L_{\rm dust}$) to total 
bolometric luminosity ($L_{\rm dust}+L_{\rm stars}$) versus Hubble type. The
filled symbols denote cluster core galaxies and the open symbols galaxies
from the cluster periphery.}
\end{figure}

For the NIR output we used the $H$ and $K^{\prime}$ magnitudes 
from Boselli et al. (1997). The $K^{\prime}$ magnitudes were available for all galaxies in
the sample, while the $H$ ones only for a few cases. Corrections for Galactic
extinction were applied
using the same procedure as in the
optical bands. For the conversion factor between magnitudes and flux densities
we used the values from Longair (1992). For the $K^{\prime}$ magnitudes we had to
interpolate to the appropriate 2.1\,${\mu}$m wavelength.

In the UV we used the derived total magnitudes at 2500\,$\AA$
taken from Rifatto, Longo \& Capaccioli (1995) and the 1650\,$\AA$ flux
densities from Deharveng et al. (1994). The observed UV luminosity was derived
by integrating the estimated UV SED until
912\,\AA. For 7 galaxies with no UV data we
extrapolated the UV SED, by using the $U-UV$ median colours derived (as a
function of Hubble type) from the 21 galaxies in our sample which have UV data. 
Since the average contribution of the observed UV to the total bolometric luminosity was 
$\sim 1\%$ for the Virgo early spirals and $\sim 8\%$ for the Virgo later spirals, the
inaccuracies incurred by this procedure are not severe.

The UV/optical/NIR SED obtained in this way for each galaxy was 
integrated to derive the observed stellar luminosity. 

\section{The ratio of the dust luminosity to the bolometric luminosity}

For the 28 ISOPHOT Virgo galaxies that have the SV96 photometry we can
derive the percentage of the dust radiative output to the total bolometric output. The 
dust luminosities were derived by augmenting the total FIR luminosity ($L_{\rm
FIR}$) with the mid-infrared (MIR) output. Since most of the
galaxies from our sample were not detected by IRAS at 12 and 25\,${\mu}$m we
made an average correction of 30$\%$ for the MIR luminosity (this was derived
from the galaxies with MIR data). The results are
plotted in Fig.~2 against Hubble type. This time a good correlation
can be seen, with the early-spirals having less stellar light re-radiated by
the dust and with the later-types having a larger contribution from the
FIR. This can be also seen from Table~2, where the mean value for the early
spirals is 15$\%$ as compared with the 30$\%$ for the later spirals. The
validity of the correlation with Hubble type was checked for any possible
alteration due to environmental effects. In Fig.~2 this is done by
plotting the cluster core and periphery galaxies with different
symbols.\footnote{The selection procedure for cluster core and periphery
galaxies is described by Tuffs et al. (2002) and is based on the cluster 
membership classification of Binggeli, Popescu \& Tammann (1993).}  Within the
available statistics no environmental dependence is found.

\begin{table}
\caption{The percentage of the total FIR luminosity to the
total bolometric luminosity (median, mean and standard deviation of the mean)}
\begin{tabular}{cccc}
\hline\hline
Type   & \multicolumn{3}{c}{$corr$} \\
       & median & mean & ${\epsilon}$ \\
\hline
S0a-Sa & 0.11   & 0.15 & 0.05 \\
Sab-Sc & 0.23   & 0.23 & 0.03 \\
Scd-Sm & 0.27   & 0.30 & 0.04 \\
\hline
\end{tabular}
\end{table}

The mean value of $L_{\rm dust}/(L_{\rm dust}+L_{\rm stars})$ for the
later spirals is the same as the canonical value of 30$\%$ obtained for the 
IRAS BGS
by Soifer \& Neugebauer (1991). This is probably due to the fact that there 
are two factors influencing this percentage, and working in opposite 
directions. The addition 
of the ISO cold dust luminosity increases the FIR contribution to the total
bolometrics. But our sample consists of more quiescent galaxies than those
from  BGS and we expect them to have smaller FIR
contributions. To check this assumption we derive the same percentage based on
our calculated $L_{60;100}^{\rm extrapolated}$, which is the ISO equivalent of
what IRAS would have seen for our sample. The resulting median percentage is now
18$\%$ for the later spirals. So indeed IRAS would have
derived a smaller percentage for our Virgo sample than for the BGS. 
By the same token, it is probable that
the contribution of dust emission to the
total luminosity of the BGS galaxies will be greater than the 30$\%$
derived from IRAS.

\begin{figure}
\plotfiddle{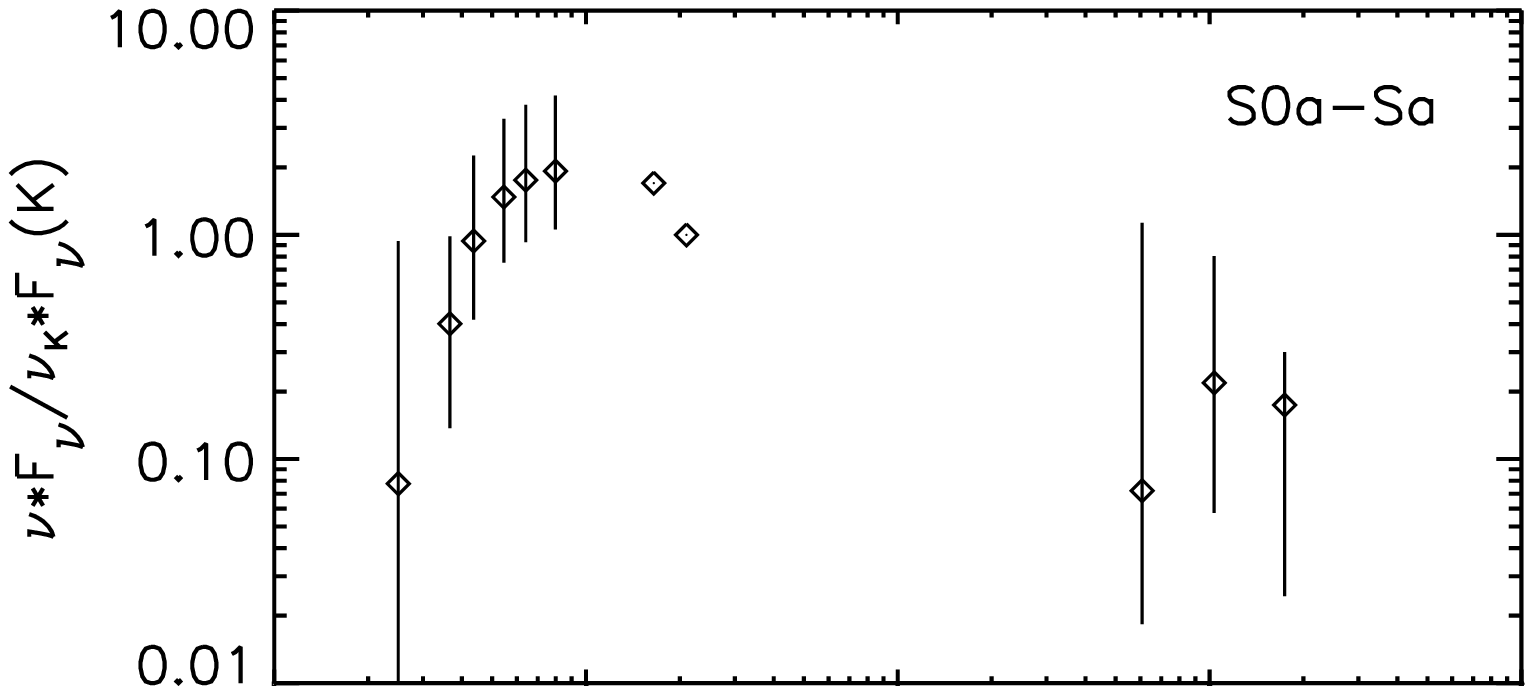}{0.0in}{0.}{50.}{50.}{-150}{-305}
\plotfiddle{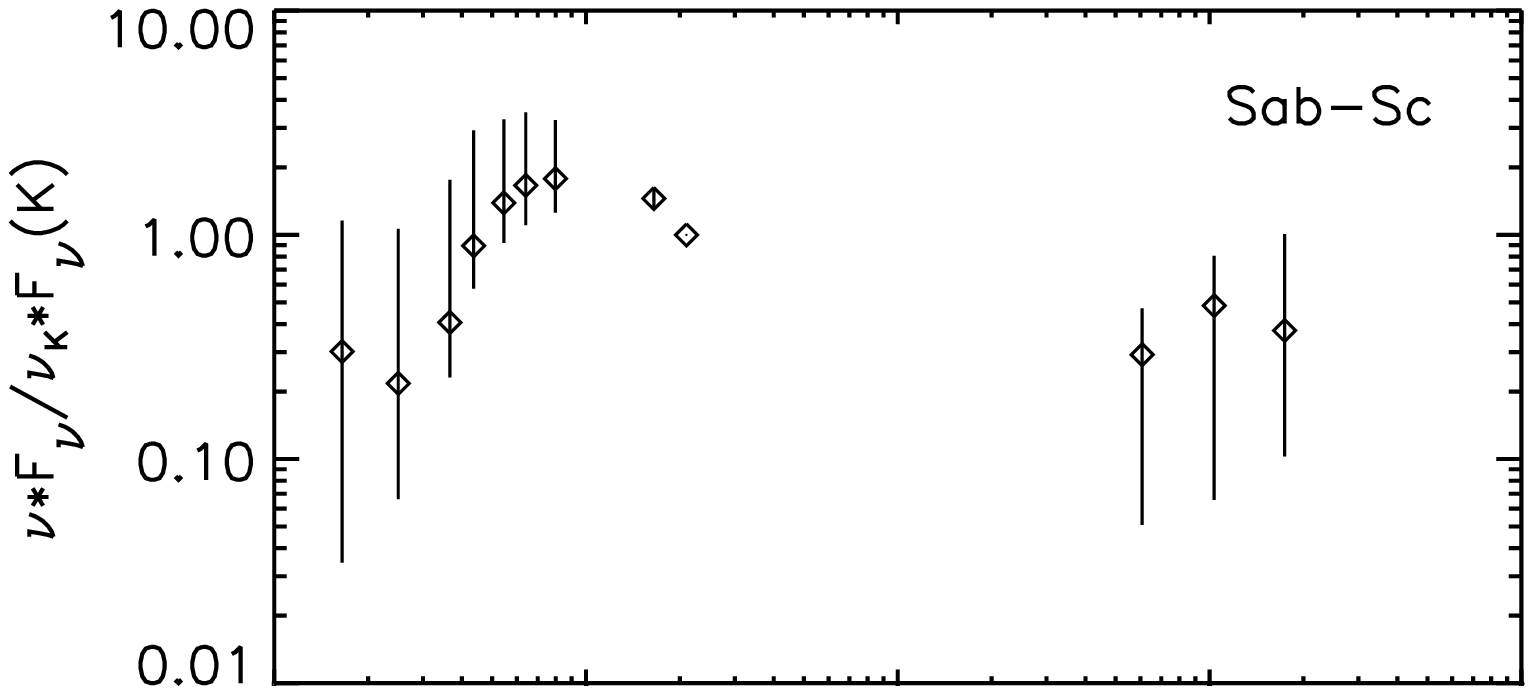}{2.2in}{0.}{50.}{50.}{-150}{-235}
\plotfiddle{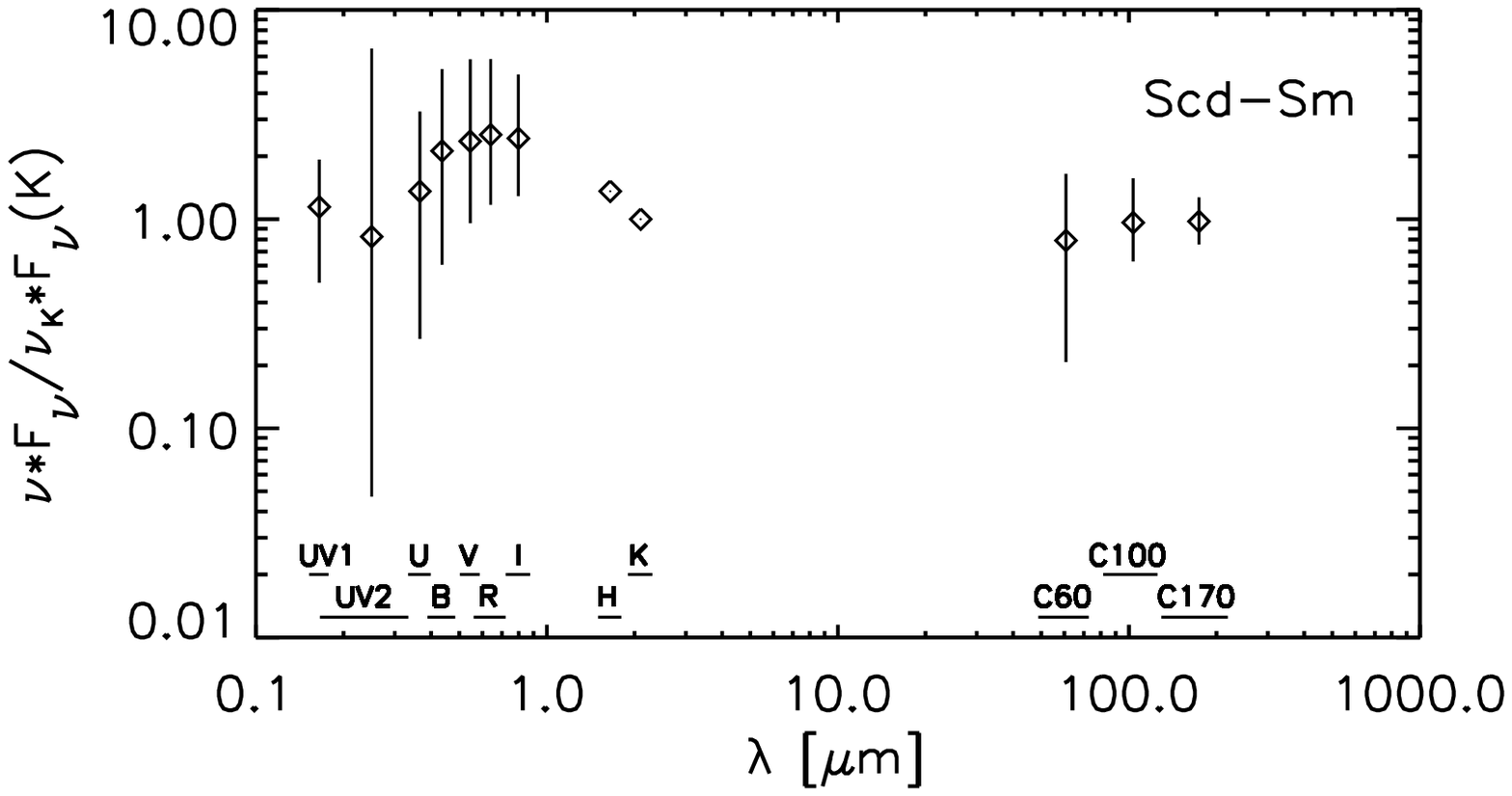}{2.2in}{0.}{50.}{50.}{-150}{-165}
\caption{Averaged SEDs (as observed), normalised to the flux in K'-band,
for subsamples of 6, 10, and 12 Virgo Cluster galaxies of type S0a-Sa, Sab-Sc and
Scd-Sm, respectively. The vertical bars denote the range of normalised
fluxes in each band. The spectral widths of the pass bands are delineated by
the horizontal bars in the lower panel. UV1 and UV2 denote bands centered on
1650 and 2500\,\AA, respectively}
\end{figure}

The percentage of stellar luminosity transformed into FIR luminosity takes 
large values for some Scd/Sm galaxies, up to 50$\%$. Furthermore, some of our 
Virgo BCDs exhibit even higher percentages (Popescu et al. 2002), increasing 
the strength of the correlation with morphological type. The energy balance 
between observed UV/optical/NIR and FIR is thus a strong function of the 
morphological type and 
this can be also seen from Fig.~3, where the averaged SEDs (as observed, not 
corrected for internal extinction) are plotted for our main bins in Hubble 
type. A general flattening can be seen when progressing from the early to the 
later types; both between the FIR and UV-optical emissions, and within the 
UV-optical regime.

No dependence of 
$L_{\rm dust}/(L_{\rm dust}+L_{\rm stars}$) on
$K^{\prime}$ magnitudes was found, indicating that the conversion of starlight 
into IR thermal light in the local universe is statistically independent of 
galaxy mass.

\section{Discussion}

We have shown that the luminosity correction for cold dust does
not correlate with morphological type and exhibits a huge
scatter, with values ranging from 0 to 150$\%$. Only the S0a/Sa galaxies have 
relatively small corrections, as many of them are lacking the two dust 
temperature components. 
The big scatter in the correction factors for cold dust
suggests that the FIR SEDs are strongly influenced by opacity effects, 
and not only by variations in the intensity and colour of the 
radiated starlight. Opacity depends on the geometrical 
distributions of the stellar populations and of the dust, 
both on large and small scales, as well as on metallicity, 
and the sheer sizes of the disks. All these quantities 
vary not only along the Hubble sequence, but also within 
a given morphological class, so that the scatter in Fig.~1 
is not surprising, and no simple recipe can be  
considered to predict the luminosity of the cold 
dust emission component. Ultimately, radiative transfer calculations have to be 
performed for the derivation of the energy densities that heat the grains and 
thus produce FIR emission (e.g. Silva et al. 1998, Popescu et al. 2000a,
Misiriotis et al. 2001, Popescu \& Tuffs 2002). Our results also imply that
there is not a one to one conversion of the 
FIR luminosity into star-formation rates. 

A second result of this paper is the correlation of the ratio of the dust to 
the stellar output with morphological type. This correlation can be also 
interpreted as a 
sequence from normal to dwarf gas rich galaxies, with the dwarfs having an 
increased contribution of the FIR output to the total bolometric output. These
findings could be important for our perception of the distant Universe, where,
according to the hierarchical galaxy formation scenarios, gas rich dwarf 
galaxies should prevail at those epochs. We would then expect these galaxies 
to 
make a higher contribution to the total FIR output in the early Universe than
previously expected. 
This, together with the cosmic-ray driven winds, in which grains can survive and be
inserted in the surrounding intergalactic medium (Popescu et al. 2000b), could
potentially change our view of the high redshifted Universe.

\section*{Acknowledgments}
We would like to thank Heinrich V\"olk for helpful discussions and
comments.

\end{document}